\documentclass[aps,prb,twocolumn,showpacs,preprintnumbers,amsmath,amssymb]{revtex4}

\usepackage[dvips,final]{graphicx}
\usepackage{dcolumn}
\usepackage{bm}
\usepackage[usenames]{color}

\usepackage{amsmath}	

\newcommand{\bra}[1]{\left<#1\right|}
\newcommand{\ket}[1]{\left|#1\right>}

\begin{document}


\title{Numerical Study of Photo-Induced Dynamics in Double-Exchange Model}
\author{Y.~Kanamori$^1$, H.~Matsueda$^{2}$, and S. Ishihara$^{1, 3}$} 
\address{$^1$Department of Physics, Tohoku University, Sendai 980-8578, Japan}
\address{$^2$Sendai National College of Technology, Sendai, 989-3128, Japan}
\address{$^3$Core Research for Evolutional Science and Technology (CREST), Tsukuba 305-0047, Japan}

\date{\today}
\begin{abstract}
Photo-induced spin and charge dynamics in double-exchange model are numerically studied.  
The Lanczos method and the density-matrix renormalization-group method are applied to one-dimensional finite-size clusters. 
By photon irradiation in a charge ordered (CO) insulator associated with antiferromagnetic (AFM) correlation, both the CO and AFM correlations collapse rapidly, and appearances of new peaks inside of an insulating gap are observed in the optical spectra and the one-particle excitation spectra. 
Time evolutions of the spin correlation and the in-gap state are correlated with each other, and are governed by the transfer integral of conduction electrons. 
Results are interpreted by the charge kink/anti-kink picture and their effective motions which depend on the localized spin correlation. 
Pump-photon density dependence of spin and charge dynamics are also studied. 
Roles of spin degree of freedom are remarkable in a case of weak photon density. 
Implications of the numerical results for the pump-probe experiments in perovskite manganites are discussed. 
\end{abstract}

\maketitle

\section{introduction}
Coherent controlling of electronic and structural phases by shining light has been one of the attractive themes not only in fundamental condensed-matter physics but also in electric and communication technologies. 
Drastic state-change by photon irradiation, often termed photo-induced phase transition, is ubiquitously seen in a variety of materials.~\cite{Nasu_PIPT,Iwai_JPSJ75} 
In particular, a number of studies in photo-irradiation effects have been done in correlated electron systems, such as transition-metal oxides, low-dimensional organic salts and others. 
This is because i) cooperative and ultra-fast changes of states by light occur due to strong electron correlation, and ii) coupling between multiple degrees of freedom of electron, i.e. spin, charge and orbital, exhibits a variety of photo-induced phenomena. 
In recent rapid progresses of experimental techniques, transient and non-equilibrium states induced by light irradiation have been examined by several kinds of time-resolved scattering and spectroscopy experiments, for example photoemission spectroscopy,\cite{Carpene_RSI80} x-ray diffraction,~\cite{Cavalleri_PRL87,Collet_Sci280} electron diffraction\cite{Carbone_PRL100} and so on in addition to the conventional optical pump-probe spectroscopy. 

Perovskite manganites $R_{1-x}$$A_x$MnO$_3$ ($R$: a rare-earth ion, $A$: an alkaline-earth ion) is one of the well studied correlated electron materials in a viewpoint of ultra-fast photo-induced phenomena. 
A nominal valence of a Mn ion is $3+x$ and the electron configuration of the $3d$ orbitals is $d^{4-x}$. One of the doubly degenerate $e_g$ orbitals is occupied by $1-x$ electron and 
the $t_{2g}$ orbitals are occupied by three electrons with parallel spin.  
One remarkable property in manganite is strong phase competition. 
Around $x=0.5$, a subtle energy balance controls stability of two electronic phases; 
charge ordered (CO) insulator associated with antiferromagnetic (AFM) order and ferromagnetic (FM) metal.~\cite{Tomioka_PRB66} 
Colossal magnetoresistance (CMR) effect is one of the examples of drastic phase change between the two phases by applying external field.~\cite{Tokura_PRL76} 
Electronic and structural phase controlling is also performed by light.~\cite{Miyano_PRL78,Fiebig_Sci280,Ogasawara_JPSJ71,McGill_PRL93,Miyasaka_PRB74,Okimoto_JPSJ76,Matsubara_PRL99,Mertelj_EPL86,Matsubara_PRB77,Matsubara_JPSJ78,Matsuzaki_PRB79} 
By photo-irradiation in CO insulating phase, the optical absorption spectra 
around the charge-transfer excitation shift to lower energy.~\cite{Ogasawara_JPSJ71}  
After 10ns, the spectra are almost recovered. These experimental results imply a generation of a transient metallic state by photo-irradiation and its fast relaxation within 10ns. The transient magnetic property was examined by the magneto-optical Kerr spectroscopy. 
A Kerr rotation appears by photo irradiation and its angle gradually increases within 1ps.~\cite{Miyasaka_PRB74,Matsubara_PRL99,Matsubara_JPSJ78} That is to say, spin and charge structures are changed cooperatively by pump photon irradiation. 
The excitation density dependence of the photo-induced state was also examined in the pump-probe experiments.~\cite{Matsubara_PRB77} The results provide a clue to resolve a stability of the photo-induced FM metal and its relaxation mechanism. 

Theoretical study of manganites has been done mainly from the view point of origin of CMR. 
The double-exchange (DE) model is not only studied as a model for manganites, but also recognized to be a standard theoretical model in solid state physics.~\cite{Yunoki_PRL80,Okamoti_PRB61,Motome_PRL91,Satoh_JMMM310} 
In this model, conduction electrons corresponding to the $e_g$ electrons couple ferromagnetically with localized spins for the $t_{2g}$ spins. 
On the contrary, transient and non-equilibrium properties in the DE
model have not been examined well so far. 

In this paper, motivated from the optical pump-probe experiments in perovskite maganites, we present a numerical study of the photo-induced dynamics in the one-dimensional DE model. 
Time dependences of the static and dynamical quantities are calculated by applying the Lanczos method and the density-matrix renormalization-group (DMRG) method in finite size clusters. 
By photo-irradiation in the CO and AFM insulating phase, both the CO and AFM correlations rapidly collapse.
New peak structures appear inside of the gap in the optical absorption spectra and the one-particle excitation spectra, and grow up with time. 
It is shown that time dependences of the spin correlation and the new peak structures 
are scaled by a universal curve. 
This implies strong coupling between the charge and spin sectors in transient photo-excited state. 
We also study the pump-photon density dependence of the photo-induced dynamics. 
Roles of spin degree are remarkable in a region of weak photon density in comparison with charge degree. Implications of the present numerical calculations for the photo-induced phenomena in manganites are discussed. 

In Sect.~\ref{sect:model}, the extended DE model and a formulation of the photo-excited states are presented. The numerical methods applied to one-dimensional clusters are introduced in Sect.~\ref{sect:method}. The ground state properties before photo-irradiation are briefly introduced in Sect.~\ref{sect:gs}. 
The main part in this paper is Sect.~\ref{sect:dynamics} where 
the numerical results for the photo-induced spin and charge dynamics are presented. The pump-photon density dependences of the photo-induced dynamics are shown in Sect.~\ref{sect:amp}. Section~\ref{sect:dis} is devoted to discussion and concluding remarks. 
Details of an effective model and analytical results in the spin-less $V-t$ model are given in Appendix A.  
A part of this paper was briefly presented in Ref.~\onlinecite{Kanamori_PRL}. 

\section{model and formulation}
\label{sect:model}

We introduce the one-dimensional extended DE model to examine both the CO insulating state associated with the AFM order and the FM metallic state.
A model Hamiltonian is given by
\begin{align}
 {\cal H}_0 &=-\alpha t \sum _{\left<ij\right> \sigma}
 \left ( c_{i \sigma}^\dag c_{j \sigma} + H.c. \right)
 +U\sum _{i} n_{i\uparrow} n_{i\downarrow} 
 \nonumber \\
 &+V \sum _{\left<ij\right>}n_{i}n_{j} 
 -J_H \sum _{i} {\bm S}_i\cdot {\bm s}_i
+J_S\sum _{\left<ij\right>} {\bm S}_i \cdot {\bm S}_{j}  , 
\label{171535_14Jan09}
\end{align}
where $c_{i \sigma}$ is the annihilation operator for a conduction electron with spin $\sigma(= \uparrow, \downarrow)$ at site $i$, and ${\bm S}_i$ is a spin operator for the localized spin.
We introduce the number operator $n_i = \sum _\sigma n_{i \sigma}= \sum _\sigma c_{i \sigma}^\dag c_{i \sigma}$ and the spin operator ${\bm s}_i=(1/2)\sum_{\alpha \beta} c_{i \alpha}^\dag \left( {\bm \sigma }\right)_{\alpha  \beta } 
c_{i \beta }$ with the Pauli matrices ${\bm \sigma }$ for the conduction electrons. 
The first term in Eq.~(\ref{171535_14Jan09}) is for the electron hopping with amplitude $\alpha t$ between the nearest neighboring (NN) sites.
The second and third terms represent the on-site Coulomb interaction $U$, and the NN Coulomb interaction $V$ for the conduction electrons, respectively.
We consider the Hund coupling $J_H(>0)$ between the conduction electron and the localized spin, and the AFM superexchange interaction $J_S(>0)$ between the NN localized spins.
We take $t=1$ as a unit of energy, and a magnitude of the transfer integral is changed by changing the parameter $\alpha$.
For simplicity, we assume a single orbital for a conduction electron and $S=1/2$ for a localized spin. 

The vector potential for the pump photon at time $\tau$ is given as ${\bm A}_{\rm pump}(\tau )=A_{\rm pump}(\tau) \hat z$ where the $z$ axis is taken to be parallel to the chain direction. 
The interaction between the conduction electrons and the pump photon is introduced as the Peirles phase: 
\begin{align}
 {\cal H'}(\tau) &= -\alpha t\sum _{\left< ij \right> \sigma} c_{i \sigma}^\dag c_{j \sigma} 
 \left [ e^{-i\int {\bm A}_{\rm pump}(\tau)  \cdot d{\bm r}}-1 \right ]  +H.c. , 
\label{eq:ha}
\end{align}
where a subtraction in the parenthesis corresponds to the first term in Eq.~(\ref{171535_14Jan09}). 
In a case of weak pump-photon density, this Hamiltonian is reduced into the following form, 
\begin{align}
 {\cal H'}(\tau)=  -j A_{\rm pump}(\tau),
\end{align}
up to the order of $O(A_{\rm pump})$. 
We introduce the current operator defined by 
$j =i\alpha t\sum_{\left< i j \right> \sigma }\left(  c_{i \sigma }^\dag
c_{j \sigma } - H.c. \right) $. 
By using the first-order time-dependent perturbation theory with respect to ${\cal H}'({\tau})$, the wave function at
time $\tau$ is obtained as
\begin{align}
\ket{\phi_0 (\tau) }=e^{-iE_0 \tau}\ket{0}+\sum _{n\neq 0} c_n(\tau) \ket{n} , 
\label{171603_25Mar09}
\end{align}
where a coefficient is given by 
\begin{align}
c_n (\tau)=i  e^{-iE_n \tau}  \bra{n}j\ket{0} 
\int ^{\tau} _{-\infty }d\tau ' A_{\rm pump}(\tau ')
 e^{-i(E_0-E_n) \tau '} ,
\label{164824_25Mar09}
\end{align}
and $\ket{n}$ is the $n$-th eigen-state of ${\cal H}_0$ with the energy $E_n$.
We assume that the system is in the ground state $\ket{0}$ at $\tau =-\infty$, and focus on the second term in Eq.~(\ref{171603_25Mar09}) which represents the one-photon absorbed state. 
From now on, this is termed $|\phi(\tau) \rangle$ as 
\begin{align}
\ket{\phi (\tau) } \equiv \sum _{n\neq 0} c_n(\tau) \ket{n} . 
\end{align} 
The vector potential for the pump-photon is assumed to be 
a damped oscillator form as 
\begin{equation}
A_{\rm pump} (\tau)=A_{\rm 1} \exp\left( -\gamma _0 |\tau| 
 -i\omega _0 \tau \right) , 
\label{eq:Apump}
\end{equation}
where $\omega_0$ is a frequency and $\gamma_0$ is a damping factor. 
We impose a condition of $\tau \gg \gamma _0 ^{-1}$ which implies that the pump photon is fully damped at time $\tau$ of our interest. 
Then, the coefficient in Eq.~(\ref{164824_25Mar09}) is obtained as 
\begin{align}
c_n(\tau) &=i A_{{1}}  e^{-iE_n \tau}
\bra{n}j\ket{0} \int ^\infty _{-\infty }d\tau '  e^{-i(\omega _0 -E_n+E_0)
 \tau '-\gamma _0 |\tau '|}  \nonumber\\
& =-2iA_{{1}}  e^{-iE_n \tau}
\bra{n}j\ket{0}  {\rm Im}\left[ \frac{1}{\omega _0
 -E_n+E_0 +i\gamma _0}\right]  .  
\label{184611_11Apr09}
\end{align}
By inserting this expression, we have the wave function at time $\tau$ as 
\begin{align}
\ket{\phi(\tau)} 
 &=-2iA_{1}\sum _{n\neq 0}
e^{-iE_n \tau} 
\nonumber \\
&\times 
 {\rm Im}\left[  \frac{1}{\omega _0
 -E_n+E_0 +i\gamma _0}\right] \ket{n}  
\bra{n}j\ket{0} 
 \nonumber\\
 &=-2iA_{1}
e^{-i{\cal H}_0 \tau}
 {\rm Im}\left[ \frac{1}{\omega _0
 -{\cal H}_0+E_0 +i\gamma _0}\right]j\ket{0},
\end{align}
where $\bra{0}j\ket{0}=0$ is used. 
Here we redefine the one-photon absorbed state by 
\begin{align}
\ket{\phi(\tau)}=\frac{1}{{\cal N}}{\rm e}^{-i{\cal H}_{0} \tau }{\rm
 Im}\left[ \frac{1}{ \omega _0 - {\cal H}_0 +E_0 +i\gamma _0 } \right] j  \
 \ket{0} ,\label{215649_17Oct08}
\end{align}
with a normalization factor ${\cal N}$.

\begin{figure}[]
\begin{center}
\includegraphics[width=\columnwidth,clip]{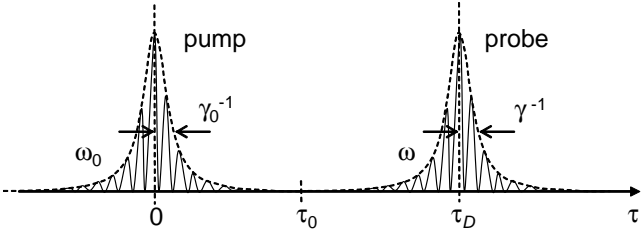}
\end{center}
\caption{
A schematic view of the wave packets for the pump and probe light intensities as a function of time. 
}
\label{fig:time}
\end{figure}
Photo-excited states are monitored by calculating the several kinds of static and dynamical quantities. 
The transient excitation spectra are formulated based on the liner-response theory.
We explain, as an example, a formulation of the transient optical absorption spectra.
The vector potential for the probe photon is set to be parallel to the $z$ direction and has a damped oscillator form as 
${\bm A}_{\rm
probe} (\tau )=A_{\rm probe}(\tau) \hat z$ 
and  
\begin{equation}
A_{\rm probe}(\tau)=A_{2} \exp \left( -\gamma |\tau-\tau _D|  -i\omega  \tau \right), 
\end{equation}
with a frequency $\omega$, the center of an envelope $\tau_D$, and a damping factor $\gamma$. 
We assume that the pump photon is fully damped at time when the probe photon comes in, and there are no interference between the pump and probe photons. 
This is given by the condition  
$\left( \tau_D - \gamma^{-1}  \right)\gg \gamma_0^{-1}$. 
A schematic picture for the pump-photon and probe-photon wave packets is shown in Fig.~\ref{fig:time}. 
The interaction Hamiltonian between the probe photon and the conduction electrons 
at time $\tau$ is given by 
\begin{align}
 {\cal H}''(\tau )= -j A_{\rm probe}(\tau) , 
\end{align}
within the first order of $A_{\rm probe}(\tau)$. 
When the probe photon is taken into account, 
the electronic wave function at time $\tau $ $(\gg \gamma _0 ^{-1}) $ is given by 
\begin{align}
 \ket {\psi (\tau)} = U(\tau ,\tau _0)\ket{\phi(\tau _0)}.
\end{align}
Here, $\ket{\phi(\tau _0)}$ is 
the wave function at time $\tau_0$ where a condition $\tau_D-\gamma^{-1} \gg \tau_0 \gg \gamma_0^{-1}$ 
is satisfied and is given by Eq.~(\ref{215649_17Oct08}). 
The time-evolution operator is defined by 
\begin{align}
 U(\tau ,\tau _0)=U_0(\tau ,\tau _0) U_1(\tau ,\tau _0),
\end{align}
 with 
\begin{align}
U_0(\tau ,\tau _0)=\exp \left [-i{\cal H}_0 \left( \tau-\tau _0 \right) \right] ,
\end{align}
and 
\begin{align}
U_1(\tau ,\tau _0)=P \exp \left[ i\int _{\tau _0} ^\tau
  d\tau 'j(\tau ',\tau _0)A_{\rm probe}(\tau')\right] ,
\end{align}
where $P$ is the time-ordered operator, and $j(\tau',\tau _0)\left[
=U_0 ^\dag (\tau ',\tau _0)jU_0(\tau ',\tau _0) \right]$ is the interaction representation of the current operator.
The expectation value of the current operator at time
$\tau$ is given as 
\begin{align}
\left< j\right>(\tau )& \equiv \left< \psi (\tau) |\ j\ | \psi(\tau) \right> \nonumber \\
&=-i\int _{\tau_0} ^{\tau} d\tau ' \bra{\phi(\tau _0)} [j(\tau',\tau_0),j(\tau,\tau_0)]
\ket{\phi(\tau _0)} A_{\rm probe} (\tau ')  
\nonumber \\
& +{\cal O}(A_{\rm probe}^2) . 
\label{182907_11Apr09}
\end{align}
Here we use a relation $\bra {\phi(\tau_0)} j(\tau , \tau_0) \ket{\phi(\tau_0)}=0$,  
because $\ket {\phi(\tau)}$ is an eigen state of parity. 
By inserting the complete set into the integrand  
and integrating out $\tau'$ under the condition 
of $(\tau _D - \tau _0) \gg \gamma ^{-1}$, 
we obtain 
\begin{align}
\left< j\right>(\tau )= \chi (\omega ,\tau) A_{\rm probe} (\tau ) +
 \chi_0( \omega, \tau) A_{\rm probe}(\tau _D) ,
\label{eq:jt}
 \end{align}
with 
 \begin{align}
& \chi (\omega ,\tau)=- \sum _{n m l} {\bar c}^\ast _n(\tau)
{\bar c}_l(\tau)j_{nm}j_{ml}
\nonumber \\
& \times \biggl [\frac{1}{\omega - E_m +E_l \pm i \gamma}
 - \frac{1}{\omega - E_n +E_m \pm i \gamma} \biggr],
\label{194249_1Apr09}
\end{align}
and 
\begin{align}
& \chi_0(\omega, \tau)= -\theta (\tau -\tau _D) 2i\gamma \sum _{n m l} j_{nm}j_{ml}
\nonumber \\
& \times  \Bigl[ 
\frac{ {\bar c}_n ^{\ast} (\tau ) {\bar c}_l (\tau _D)e^{-iE_m (\tau -\tau _D)}}{(\omega -E_m +E_l)^2+\gamma ^2} 
-\frac{ {\bar c}_n ^\ast (\tau _D) {\bar c}_l (\tau )e^{iE_m (\tau -\tau _D)}}{(\omega -E_n +E_m)^2+\gamma ^2} 
\Bigr]. 
\end{align}
We introduce ${\bar c}_n (\tau)=\left<n|\phi(\tau )\right>$ which is proportional to $c_n(\tau)$ in Eq.~(\ref{184611_11Apr09}), and $j_{nm}=\bra{n}j\ket{m}$.
The first and second terms in Eq.~(\ref{eq:jt}) represent the induced current which is proportional to the probe-photon vector potential $A_{\rm probe}(\tau)$, 
and the oscillating part which remains to be finite at $\tau= +\infty$, respectively.  
The positive and negative signs of $i\gamma$ in the denominators of Eq.~(\ref{194249_1Apr09}) are for the cases of $\tau < \tau _D$ and $\tau > \tau _D$, respectively. 
We focus on the first term in Eq.~(\ref{eq:jt}) with the positive sign of $i \gamma$ in $\chi(\omega, \tau)$, from now on. 
The first and second terms in $\chi(\omega, \tau)$ represents the photon-absorption and photon-emission processes, respectively. 
The optical absorption spectra is defined by the imaginary part of the first term given as 
 \begin{align}
& \alpha (\omega ,\tau)= -\frac{1}{L\pi}{\rm Im} \sum _{n m l}
  \frac{{\bar c}^\ast _n(\tau){\bar c}_l(\tau)j_{nm}j_{ml}}{\omega - E_m +E_l + i
  \gamma},
  \label{eq:alpha}
\end{align}
where $L$ is the system size.

In the similar way, the one-particle excitation spectra are obtained as a sum of the electron and hole parts:  
$A(q, \omega)=A^{\rm hole}(q, \omega)+A^{\rm ele}(q, \omega)$ with 
\begin{align}
&A^{\rm hole} (q,\omega )=\sum _{n m l \sigma } {\rm Im}
\frac{{\bar c}^\ast _n(\tau){\bar c}_l(\tau)
\bra{n} c_{q \sigma} \ket{m} \bra{m} c_{-q \sigma} ^\dag \ket{l}}{-\pi(\omega - E_m +E_l + i \gamma)},
\end{align}
and 
\begin{align}
A^{\rm ele} (q,\omega )=\sum _{n m l \sigma } {\rm Im}
\frac{{\bar c}^\ast _n(\tau){\bar c}_l(\tau)
\bra{n}c_{q \sigma} ^\dag \ket{m} \bra{m}c_{-q \sigma}\ket{l}}{-\pi(-\omega - E_m +E_l + i \gamma)}.
\end{align}
We also obtain the imaginary part of the dynamical charge susceptibility as 
\begin{align}
&N  (q,\omega )=- \frac{1}{\pi}\sum _{n m l} {\bar c}^\ast _n(\tau)
{\bar c}_l(\tau)\bra{n}n_q\ket{m} \bra{m}n_{-q}\ket{l}
\nonumber \\
&\times {\rm Im}
\left[\frac{1}{\omega - E_m +E_l + i \gamma}
- \frac{1}{\omega - E_n +E_m + i \gamma} \right],\label{172707_8Apr09} 
\end{align}
 and that of the dynamical spin susceptibility for the localized spin as 
\begin{align}
&S (q,\omega )=- \frac{1}{\pi}\sum _{n m l} {\bar c}^\ast _n(\tau)
{\bar c}_l(\tau)\bra{n}{\bm S}_q \ket{m} \bra{m}{\bm S}_{-q}\ket{l}
\nonumber \\
&\times  {\rm Im}\left[\frac{1}{\omega - E_m +E_l + i \gamma}
- \frac{1}{\omega - E_n +E_m + i \gamma} \right] , 
\label{172721_8Apr09}
\end{align}
where $c_{q \sigma}$, $n_q$, and ${\bm S}_q$ are the Fourier transforms of
$c_{i \sigma}$, $n_i$, and ${\bm S}_i$, respectively.

In the numerical calculation, to reduce the computer resource, 
$E_n$ and $E_l$ in the denominator of
Eq.~(\ref{194249_1Apr09}) are replaced by 
the expectation value for ${\cal H}_0$ with respect to the wave function after the pump irradiation, i.e. $E_p=\bra{\phi(\tau)}{\cal H}_0\ket{ \phi(\tau)}$.
Then the response function is rewritten as 
\begin{align}
& \chi (\omega ,\tau)  \simeq - 
\bra{\phi (\tau)} j 
\nonumber \\ & \times \biggl[ \frac{1}{\omega - {\cal H}_0 +E_p + i \gamma}
 -\frac{1}{\omega - E_p +{\cal H}_0 + i \gamma} \biggr]j\ket{\phi (\tau)} . 
\end{align}
The same approximation is applied to the optical absorption spectra and the one-particle excitation spectra. 
This approximation is reasonable in the present case where the pump-photon energy is tuned around a narrow energy region of 
$\omega_0 \pm \gamma_0$ with $\gamma_0 \ll \omega_0$. 
We have numerically confirmed in the calculation of $\alpha(\omega)$ that this approximation reproduces quantitatively the results based on the exact expression in Eq.~(\ref{eq:alpha}).
In order to unify the expressions before and after pumping, 
we introduce the following expression for the optical absorption spectra: 
\begin{align}
\alpha (\omega )&=
-\frac{1}{L\pi} {\rm Im}\left< j 
\frac{1}{\omega - {\cal H}_0 + \left < {\cal H}_0 \right >  + i \gamma} j \right>.
\end{align}
A bracket $\left<\cdots \right>$ implies the expectation value with respect to $\ket{0}$ before pumping, and that with respect to 
$\ket{\phi(\tau)}$ after pumping.

\section{method}
\label{sect:method}
%
In order to analyze the photo-excited states in the generalized DE model, the ED method based on the Lanczos algorithm and the DMRG method~\cite{White_PRL69,White_PRB48,Hallberg_PRB52,White_PRL93,Matsueda_JPSJL} are applied into one-dimensional finite-size clusters.

In the ED method, one-photon absorbed state at $\tau=0$ 
is calculated by inserting a subset $\{ \ket{\tilde{n}} \}$ with finite dimension $M_1$, 
instead of the complete set $\{ \ket{n} \}$, in Eq.~(\ref{215649_17Oct08}) as  
\begin{align}
\ket{\phi(\tau =0)}
\simeq \frac{1}{{\cal N}}
\sum _{\tilde{n}} ^{M_1} 
{\rm  Im}\left[ \frac{\bra{\tilde{n}} j   \ket{0}}{ \omega _0 -
 E_{\tilde n} +E_0 +i\gamma _0 } \right]  \ket{\tilde{n}}
 .\label{180000_Sep06} 
\end{align} 
This set of the wave functions is obtained by the Lanczos procedure with the 
$M_1$ steps from a trial function $j\ket{0}$, 
and $E_{\tilde n}$'s are the corresponding eigen energies.  
We take $M_1=300-400$ which is enough for the optical absorption spectra before pumping. 
This reduction technique is also used to calculate the time evolution of the wave function. 
The wave function at time $\tau +\delta \tau$ is obtained from $\ket{\phi(\tau)}$~\cite{Park_JCP}
\begin{align}
 &\ket{\phi(\tau  +\delta \tau) } \simeq \sum _{\tilde{n}} ^{M_2} e^{-i E_{\tilde n} \delta \tau}
\ket{\tilde n} \left< \tilde{n} | \phi(\tau ) \right> ,
\label{184408_9Nov09}
\end{align}
where a subset with a dimension $M_2$ is obtained by the Lanczos procedure for the Hamiltonian ${\cal H}_0$ from the trial function $\ket{\phi (\tau )}$. 
We take $M_2=20$ and $ \delta \tau =10^{-2}/t- 10^{-3}/t$ 
which are enough for the several time-dependent quantities. 
We checked that the total energy remains to be a constant within a numerical error. 

In the DMRG method, we use the multi-target DMRG algorithm to obtain the excitation spectra and the time evolution. 
The density matrix for the system block is set to be 
\begin{align}
\rho _{ii'}
=\sum _{\alpha j}p^{(\alpha)} \psi  _{i j}^{(\alpha) \ast}\psi  _{i'j}^{(\alpha) } , 
\label{eq:dm}
\end{align}
where $p^{(\alpha)}$ is the weight for the $\alpha$-th target with a relation $\sum _\alpha
p^{(\alpha)}=1$, $i (i')$ and $j$ represent the states in the system and environment blocks, respectively, and $\psi^{(\alpha) }$ is the wave function
of the $\alpha$-th target state.
For the excitation spectra with respect to the operator ${\cal O}$ before pumping, 
we take the following target states: 
\begin{align}
 &\left\{ \ket{0}, {\cal O} \ket{0}, 
\frac{1}{\omega -{\cal H}_0+E_0 +i\gamma}{\cal O} \ket{0}\right\} . 
\label{131344_11Nov09} 
\end{align}
For example, we take ${\cal O}=j$ for the optical absorption spectra. 
In the calculation of the correction vector in Eq.~(\ref{131344_11Nov09}), 
we use the reduced basis set obtained by the Lanczos procedure from the trial wave function ${\cal O} \ket{0}$. 
To obtain the wave function $\ket {\phi(\tau) }$ at time $\tau$ after pumping, 
we take the following target states: 
\begin{align}
\left \{ \ket{0},
j\ket{0}, \ket{ \phi(\tau )} \right \} , 
\end{align}
where we define  
\begin{align}
 &\ket{ \phi(\tau)} \simeq \frac{1}{{\cal N}}
\sum _{\tilde{n}} ^{M_3} e^{-i E_{\tilde n} \tau}
{\rm  Im}\left[ \frac{\bra{\tilde{n}} j   \ket{0}}{ \omega _0 -
 E_{\tilde n} +E_0 +i\gamma _0 } \right] \ket{\tilde{n}} 
 .
 \label{190429_9Nov09}
\end{align}
Here, $\{ \ket{\tilde n } \}$ is the reduced basis set with dimension $M_3=300-400$ obtained by the Lanczos procedure in ${\cal H}_0$ from the trial function $j\ket{0}$, and $E_{\tilde n}$'s are the corresponding eigen energies. 
The total energy remains to be a constant within numerical errors. 
In the calculation of the transient excitation spectra at time $\tau$, the following target states are adopted,  
\begin{align}
\left \{ \ket{0},
j\ket{0}, \ket{\phi(\tau)}, {\cal O}\ket{\phi(\tau)}, 
\frac{1}{\omega -{\cal H}_0+E_p+i\gamma}
{\cal O} \ket{\phi(\tau )} \right \}  .
\end{align}
The truncation number is chosen to be $m=300$ in most of the calculations, 
and $m=800$ in some cases. 

To remove boundary effects, system size $L$ is set to be odd, and a number of the conduction electrons are $N=(L+1)/2$ which correspond to the quarter filling in the open-boundary condition. 
We take $L=9$ and 13, and confirmed no qualitative differences between the two results.
We mainly show the results in $L=9$ and $N=5$.


\section{Initial State}
\label{sect:gs}

\begin{figure}[]
\begin{center}
\includegraphics[width=\columnwidth,clip]{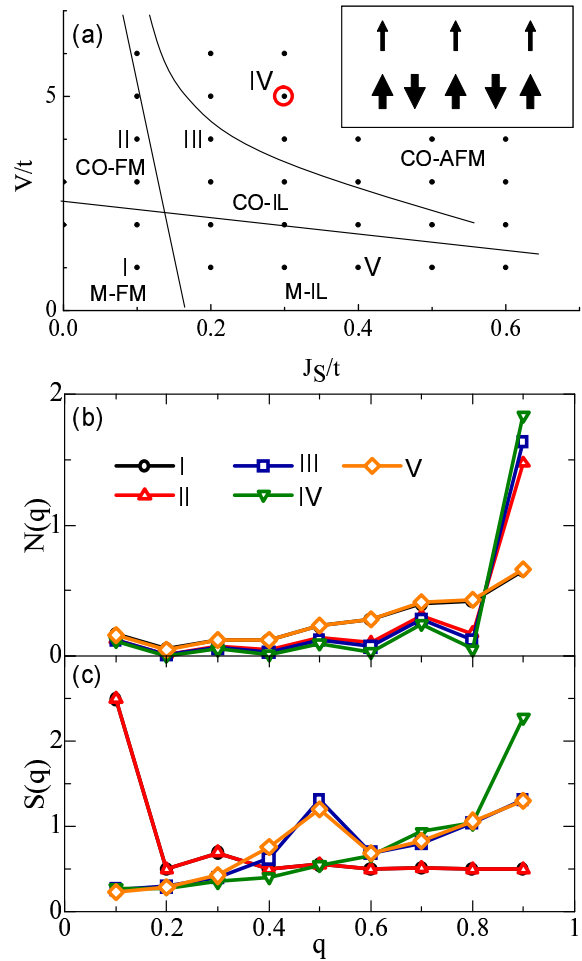}
\end{center}
\caption{
(color online)
(a) Ground-state phase diagram. Dots represent the calculated points. 
Abbreviations CO, M, FM, AFM and IL imply the charge ordered phase, the metallic phase, the ferromagnetic phase, the antiferromagnetic phase and the island phase, respectively. The inset shows a schematic picture for the spin and charge configurations in the CO-AFM phase.
A circle represents the parameters adopted in the numerical calculation
 in Sects.~\ref{sect:dynamics} and~\ref{sect:amp}. 
(b) Charge correlation function $N(q)$, and (c) spin correlation function $S(q)$ for the localized spins. 
Symbols I-V correspond to the parameters in (a). 
Numerical data of $S(q)$ in I and II are almost overlapped. 
Other parameters are chosen to be $L=9$, $U=20t$ and $J_H =15t$.
}
\label{fig:gsphase}
\end{figure}
First we show the electronic structure before photo-irradiation. 
In Fig.~\ref{fig:gsphase}(a), the phase diagram is presented in the plane of $V$ and $J_S$. 
This is obtained by the charge correlation function defined by 
\begin{align}
N(q)=\frac{2}{L+1}
\sum _{i j} \sin q r_i \sin q r_j  
\left< \delta n_i \delta n_j  \right>, 
\end{align}
with the charge fluctuation operator $\delta n_i= n_i -N/L$, and 
the spin correlation function for the localized spin defined by 
\begin{align}
S(q)=\frac{2}{L+1}\sum _{i j} \sin q r_i \sin q r_j  \left< {\bm  S}_i\cdot {\bm S}_j\right> . 
\end{align}
The correlation functions in each phase are shown in Fig.~\ref{fig:gsphase}(b) and Fig.~\ref{fig:gsphase}(c). 
Because of the open-boundary condition adopted in the present cluster, 
we introduce the quasi-momentum~\cite{Benthien_PRL} in a potential well of width $L$ defined by
\begin{align}
q=\frac{n\pi}{L+1}, 
\end{align}
with an integer $n(=1,2,\cdots ,L)$, and the momentum representation of a local operator ${\cal O}_i$ as 
\begin{align}
{\cal O}_q =\sqrt{\frac{2}{L+1}} \sum _i {\cal O}_i \sin q r_i , 
\end{align}
which takes zero at $i=0$ and $L+1$. 
It is shown in the figures that, in the CO phase, $N(q)$ takes a sharp peak at $q=\pi$. 
There are three CO phases; 1) the CO-FM phase, 
2) the CO island (CO-IL) phase where $S(q)$ has a weak peak, 
and 3) the CO-AFM phase. 
In the CO-AFM phase of our present interest, 
total spin-quantum number defined by $\vec S^{tot}=\sum_{i}
(\vec S_i + \vec s_i)$ is $(L+3)/4$, 
and a correlation function between a localized spin and a conduction electron at the same site is $N^{-1} \sum _{i} \langle {\bm S}_i\cdot {\bm s}_i
\rangle=0.247$.
These imply a ferrimagnetic configuration as shown in the inset of Fig.~\ref{fig:gsphase}(a) where the conduction electrons and localized spins form the spin-triplet states at every other sites. 
The obtained phase diagram is qualitatively consistent with the previous result in Ref.~\onlinecite{Garcia_PRB} 

The photo-induced spin and charge dynamics are examined in the CO-AFM phase near the phase boundary. 
A set of the parameter values are $U=20t, \ V=5t,\ J_H=15t,\ J_S=0.3t$ which are marked by a circle in Fig.~\ref{fig:gsphase}(a). These values are somewhat larger than the realistic values for the perovskite manganite, but are required to realize the CO-AFM phase in one-dimensional clusters. 

\begin{figure}[]
\begin{center}
\includegraphics[width=\columnwidth,clip]{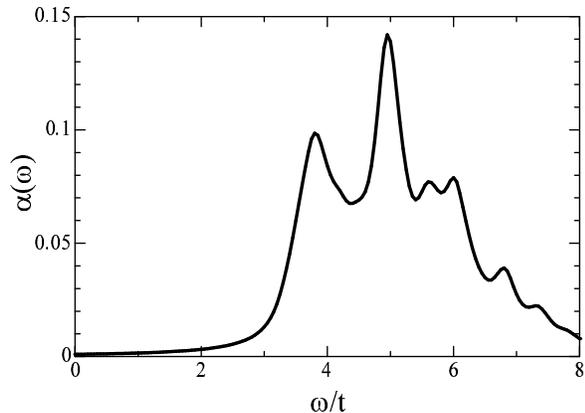}
\end{center}
\caption{
Optical absorption spectra $\alpha (\omega)$. Parameters are chosen to be $\gamma =0.2t$ and $L=9$. 
}
\label{fig:gsNSaw}
\end{figure}
The optical absorption spectra calculated in this parameter set is shown in Fig.~\ref{fig:gsNSaw}. 
In naive sense, these optical spectra correspond to the excitations from the alternate charge configuration $\cdots 1010101 \cdots $ in a chain to the configuration $  \cdots 1011001 \cdots $ 
where 1 and 0 represent the electron occupied and unoccupied sites, respectively. 
In the next section, we focus on the photo-induced dynamics where the pump-photon energy is tuned around the edge of these spectra.

\section{Photo Induced Dynamics}
\label{sect:dynamics}

In this section, we introduce the numerical results for the photo-induced dynamics. 
The pump energy is tuned at the lowest peak in $\alpha(\omega)$ 
in Fig.~\ref{fig:gsNSaw}, i.e. $\omega _0 =3.8 t$ with a damping constant $\gamma _0=0.4t$. 

\subsection{Static Correlations}
\label{subsec:static}

\begin{figure}[]
\begin{center}    
\includegraphics[width=\columnwidth,clip]{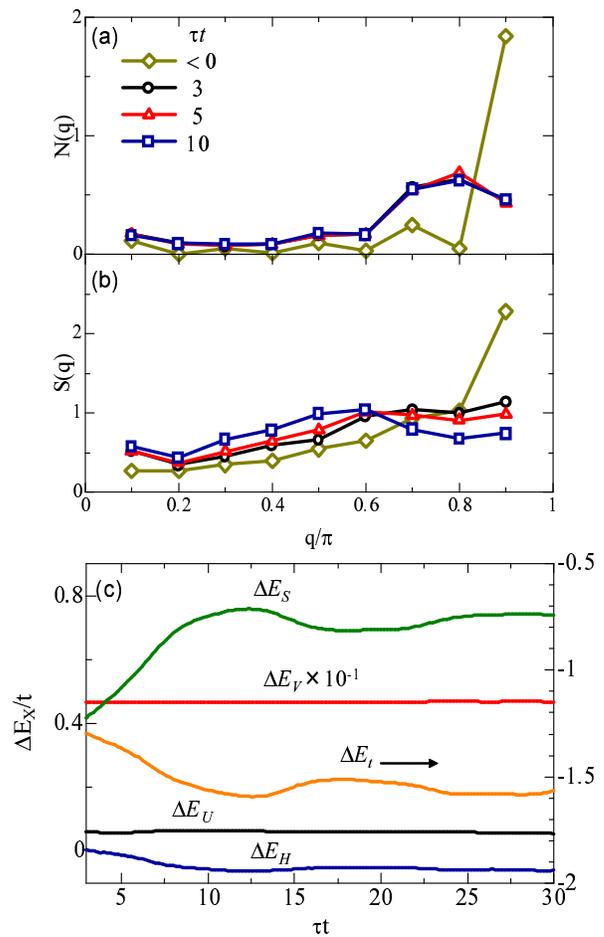}
\end{center}
\caption{
(color online) (a) Charge correlation functions, and (b) spin correlation functions of localized spins at various times. (c) Time dependence of the difference energy for each term in the Hamiltonian $\Delta E_X$.    
The NN Coulomb-interaction term is scaled by 1/10. 
}
\label{fig:NqSqHsHt_CES}
\end{figure}
%
The charge and spin correlation functions for several times are shown in Figs.~\ref{fig:NqSqHsHt_CES}(a) and \ref{fig:NqSqHsHt_CES}(b), respectively.
Just after the photo-irradiation, sharp peaks in $N(q)$ and $S(q)$ around $q=\pi$ are suppressed. After $\tau t = 3$, $N(q)$ is
almost independent of time, but $S(q)$ in large (small) $q$ regions decreases (increases) weakly with time.
We also show the time dependence of the expectation value for each term in the Hamiltonian in Eq.~(\ref{171535_14Jan09}). 
We define  
\begin{align}
E_t=  -\alpha t \sum _{\left<i j \right> \sigma}
\left \langle \left ( c_{i \sigma}^\dag c_{j \sigma}+H.c. \right ) \right \rangle , 
\label{eq:et}
\end{align}
\begin{align}
E_U = U\sum _{i} \langle n_{i \uparrow} n_{i \downarrow} \rangle,  
\end{align}
\begin{align}
E_V =  V\sum_{\left<ij\right>}
\langle n_{i}n_{j} \rangle, 
\end{align}
\begin{align}
E_H= -J_H \sum _{i} \langle {\bm S}_i\cdot {\bm s}_i \rangle , 
\end{align}
\begin{align}
E_S=  J_S\sum _{\left<ij\right>} 
\langle {\bm S}_i\cdot {\bm S}_{j} \rangle , 
\label{eq:es}
\end{align}
and differences between the expectation values at time $\tau(>0)$ and those before photo-irradiation defined by 
\begin{align}
 \Delta E_X(\tau) = E_X(\tau)-E_X^{\rm GS},
\end{align}
for $X=(t, U, V, H, S)$. 
Brackets $\langle \cdots \rangle $ in Eqs.~(\ref{eq:et})-(\ref{eq:es}) represent the expectations in terms of $|\phi(\tau) \rangle$ and $|0 \rangle $ for $E_X(\tau)$ and $E_X^{\rm GS}$, respectively.  
Results are plotted in Fig.~\ref{fig:NqSqHsHt_CES} (c).
Just after the photo-irradiation, a large increasing of the Coulomb interaction term $\Delta E_V$ implies a melting of CO. This term does not show remarkable time dependence furthermore, and the initial CO state is not restored within this time scale. 
Remarkable time dependence is also seen in the kinetic energy term $\Delta E_t$, and is almost compensated by that in $\Delta E_S$. 
That is, the energy flows from the conduction electrons to the localized spins. 
Neither remarkable change by the photo-irradiation nor large time dependence are seen in $\Delta E_H$. The local spin-triplet states between the conduction electrons and the localized spins are maintained, because the pump-photon energy is much smaller than the Hund coupling.

\subsection{Optical Absorption Spectra}
\label{subsec:opt}

\begin{figure}[]
\begin{center}
\includegraphics[width=\columnwidth,clip]{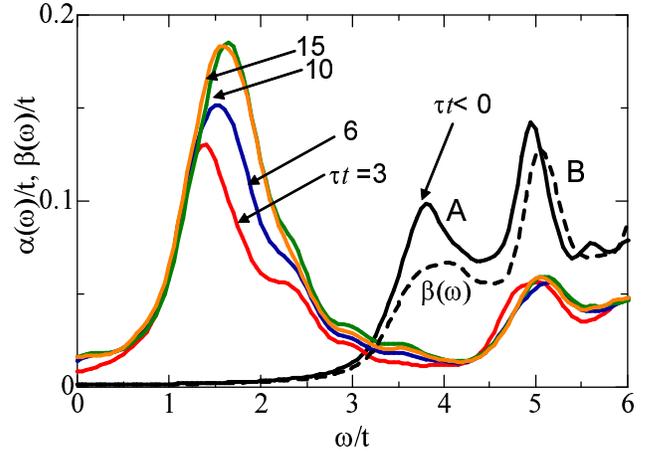}
\end{center}
\caption{(color online)
Optical absorption spectra at various times. 
Broken line is for the dynamical correlation function for the stress-tensor operator before pumping. System size is chosen to be $L=9$.
}
\label{fig:OAS_CES}
\end{figure}
\begin{figure}[]
\begin{center}
\includegraphics[width=\columnwidth,clip]{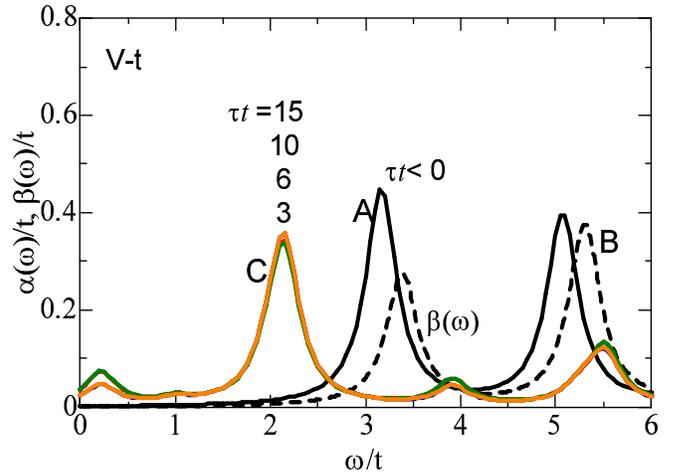}
\end{center}
\caption{(color online)
Optical absorption spectra in the spin-less $V-t$ model. 
Broken line is for the dynamical correlation function for the stress-tensor operator before pumping. Numerical data for $\tau t=$3, 6, 10, 15 are almost overlapped.   
Parameters are chosen to be $V/t=5$, $L=9$ and $N=5$.}
\label{fig:VtOAS_CES}
\end{figure}

The optical absorption spectra before and after pumping are presented in
Fig.~\ref{fig:OAS_CES}.
We also plot the dynamical correlation function for the stress-tensor operator before pumping. 
This is defined by 
\begin{align}
\beta (\omega )&=
-\frac{1}{L\pi} {\rm Im} \bra{0} \delta \eta  
\frac{1}{\omega - {\cal H}_0 +E_0 + i \gamma} \delta \eta \ket{0} ,
\end{align}
where $\delta \eta$ is the stress-tensor operator given by $\delta \eta =\eta -\bra{0}\eta \ket{0}$ and 
\begin{align}
 \eta= \alpha t\sum_{\left< ij \right> \sigma }\left(  c_{i \sigma }^\dag
c_{j \sigma } + H.c. \right) . 
\end{align}
A parity in the final states in $\beta(\omega)$ is the same with a parity in both the ground state $|0 \rangle$ and the final states in the transient-optical spectra $\alpha(\omega)$ after photo-irradiation, but is different from a parity in the one-photon absorbed state $|\phi(\tau) \rangle$ in Eq.~(\ref{215649_17Oct08}).
After the photo-irradiation, a new peak appears inside of the optical gap and grows up with time. A position of this peak around $\omega \sim 1.5t$ corresponds to 
the energy difference between the two peaks denoted by A in $\alpha
(\omega)$ and B in $\beta (\omega)$ before pumping shown in Fig.~\ref{fig:OAS_CES}.
This in-gap spectral weight is attributed to the excitations between the two, 
and corresponds to the Drude component in the thermodynamic limit as explained later. 
It is checked that a small weight around $\omega=0.2t$ vanishes in the large limit of $V/t$. 

In order to understand, the photo-induced in-gap state in $\alpha(\omega)$ in more detail, we focus on the charge degree of freedom and study the spin-less $V-t$ model in one-dimensional chain defined by 
\begin{align}
{\cal H}_{Vt}=-t \sum _{\left<ij\right>}
   \left ( d_{i}^\dag d_{j} + H.c. \right ) 
   +V\sum _{\left<ij\right>} n_i n_j . 
\label{eq:vt}
\end{align}
Here $d_i$ is the annihilation operator of a spin-less fermion at site $i$ 
and $n_i=d_i^\dagger d_i$ is the number operator.
We consider the half-filling case where the average number of fermion per site is 0.5.
The Lanczos algorithm is applied into the one-dimensional cluster with the open-boundary condition. 
The optical absorption spectra are shown in Fig.~\ref{fig:VtOAS_CES}. 
By photo-irradiation where the pump photon energy is tuned at
$\omega _0 =3.2t$ with a dumping factor $\gamma _0 =0.4t$, 
a new peak appears inside of the optical gap. This component is attributed to the excitation between the peak A in $\alpha(\omega)$ and the peak B in $\beta(\omega)$ in Fig.~\ref{fig:VtOAS_CES},  
as well as the in-gap component in the DE model shown in Fig.~\ref{fig:OAS_CES}. 
Therefore, the in-gap component shown in the DE model is attributed mainly to the charge degree of freedom and its mechanism is able to be examined within the $V-t$ model.
On the other hand, the time dependence of the in-gap band in the DE model is qualitatively different from that in the $V-t$ model where the new peak is almost time-independent after photo-irradiation. 
Spin degree of freedom in the DE model plays main roles on the time-dependence of the in-gap component. 

A charge configuration before pumping and that in the charge-excited state in the $V-t$ model are schematically represented in a one-dimensional chain as $\cdots 1010101 \cdots $ and $  \cdots 1011001 \cdots $, respectively. 
Here, 1 and 0 imply the electron occupied and unoccupied sites, respectively. 
In the photo-excited states, there are one NN electron pair represented by $11$ and one hole pair $00$ which are equivalent to a kink and an anti-kink in a chain. 
The photo-excited state is characterized by the relative momentum $q_r \equiv (Q_e-Q_h)/2$ between the momentum of the electron pair $Q_e$ and that of the hole pair $Q_h$, since the total momentum $Q_e+Q_h$ is conserved in the photo-absorption processes. 
The several peaks in $\alpha(\omega)$ shown in Fig.~\ref{fig:VtOAS_CES} are classified by $q_r$. 
Based on this kink/anti-kink picture, we derive the effective model for 
the photo-induced dynamics in the $V-t$ model in the limit of $V \gg t$. 
We consider one kink/anti-kink pair. 
Details are presented in Appendix~\ref{appendixa}. 
We obtain the transition probability 
for the current operator 
from the kink/anti-kink ground state with the relative momentum $q_{ri}=\pi/L$ to an excited state with $q_{rf}$.  
The result is given by 
\begin{align}
I \left (q_{ri}, q_{rf}  \right)=|\bra{q_{rf}}j\ket{q_{ri}}|^2={\cal C} \frac{t^2}{\pi ^2 }  
\left ( \frac{ q_{rf} \sin q_{ri} }{ q_{rf}^2-q_{ri}^2} \right )^2  ,
\label{eq:pro}
\end{align}
where a constant ${\cal C}=144$ for the lowest excited kink/anti-kink
state with the momentum  $q_{rf}=q_{ri}+\pi/L$ and ${\cal C}=64$ for other $q_{rf}$'s. 
The numerical results shown in Fig.~\ref{fig:VtOAS_CES} are well reproduced 
by this analytical method based on the kink/anti-kink picture. 
In particular, it is clarified that the lowest energy peak around $\omega \sim 2.1 t$ marked by C corresponds to the transition given in Eq.~(\ref{eq:pro}) with the minimum momentum transfer $q_{rf}-q_{ri}=\pi/L$. 
In the thermodynamic limit of $L \rightarrow \infty$, when the kink/anti-kink density $n$ is fixed, 
a position of this peak is 
$\varepsilon_{q_{ri} +\frac{\pi}{L}} -\varepsilon _{q_{ri}} \rightarrow 0$ 
and its weight is given by 
\begin{align}
D_{Vt} = \lim _{L\to \infty} nL \frac{\pi}{L}
\frac{|\bra{q_{ri}+\frac{\pi}{L}} j \ket{q_{ri}}|^2}{\varepsilon
 _{q_{ri} +\frac{\pi}{L}} -\varepsilon _{q_{ri}}}=
 nL\frac{9t}{\pi ^2}\sin q_{ri},\label{115716_2Mar10}
\end{align}
where $\varepsilon _{q}$ is the energy of the spin-less fermion.  
Here, we neglect the interaction between kinks and anti-kinks.  
This corresponds to the Drude part of the optical conductivity.  

Some aspects in $\alpha(\omega)$ in the DE model is explained from this charge kink/anti-kink picture. 
By changing the parameter values in the numerical calculation, we have confirmed that the main structures of $\alpha(\omega)$ in the DE model before pumping, e.g. the number of peaks, the $V$ dependence of the peak position and others, correspond to the structures in the $V-t$ model. 
Thus, we expect that the peaks in the DE model are classified by the relative momentum $q_r$ of the kink and anti-kink excitations, and that the photo-induced in-gap component corresponds to the Drude weight in the thermodynamic limit. 
On the other hand, the fine structures in $\alpha(\omega)$ and the time-dependence of the in-gap component in the DE model are not explained by the kink/anti-kink picture and are sensitive to the parameter $J_S$. These are attributed to a time dependence of the spin structure as explained latter.

\begin{figure}[]
\begin{center}
\includegraphics[width=\columnwidth,clip]{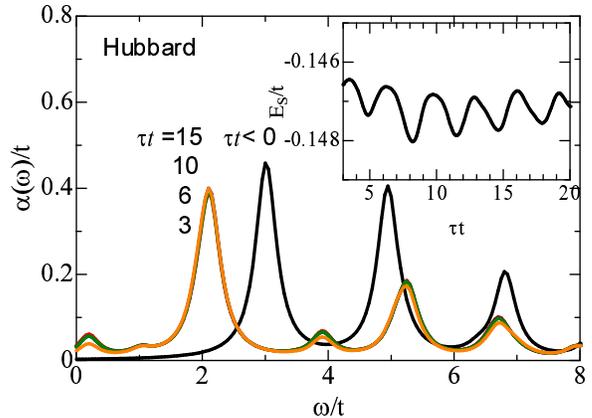} 
\end{center}
\caption{(color online) Optical absorption spectra in the extended
 Hubbard model. Numerical data for $\tau t=$3, 6, 10, 15 are almost
 overlapped. The inset is for the time dependence of
 superexchange-interaction energy defined by
 Eq.~(\ref{eq:eshubbard}). The parameters are chosen to be $U=20t$,
 $V=5t$, $L=9$ and $N=5$.}
\label{fig:Hub}
\end{figure}
As well as the $V-t$ model, almost no-time dependence of the transient-optical absorption spectra is seen in the extended Hubbard model in one-dimensional chain:  
\begin{align}
{\cal H}_{\rm Hub}&=-t \sum _{\left<ij\right> \sigma}
   \left ( c_{i \sigma}^\dag c_{j \sigma} + H.c. \right ) \nonumber \\
   &+U\sum _{i} n_{i \uparrow} n_{i \downarrow} 
   +V\sum _{\left<ij\right>}n_{i}n_{j} . 
\end{align}
The numerical results of $\alpha(\omega)$ are shown in Fig.~\ref{fig:Hub}.
We adopt the one-dimensional finite-size cluster with the open boundary condition.
The electron number per site is 0.5, and the pump photon energy and the
dumping factor are set to be $\omega _0 =3.2t$ and $\gamma _0 =0.4t$,
respectively. 
As well as the DE model and the $V-t$ model, the in-gap spectral component appears around $\omega \sim 2t$ by photo-irradiation. This component does not show remarkable time dependence. 
We also calculate the exchange-interaction energy between the NN spins defined by 
\begin{align}
E_S ^{\rm Hub} = \frac{4t^2}{U-V} \sum _{\left< ij \right> } \langle {\bm
 s}_i \cdot {\bm s}_j \rangle,
\label{eq:eshubbard}
\end{align}
with the spin operator 
${\bm s}_i=(1/2)\sum_{\alpha \beta} c_{i \alpha}^\dag {\bm \sigma }_{\alpha \beta } c_{i \beta }$. 
It is shown in the inset of Fig.~\ref{fig:Hub} that 
the change in $E^{\rm Hub}_S$ is less than 1\% of the pump photon energy.
Similar weak time-dependence in the spin sector was also observed in the half-filled Hubbard model.~\cite{Takahashi_PRL}  Through the analyses in the $V-t$ model and Hubbard model, we conclude that the localized-spin degree of freedom which couples with the conduction electrons is essential for the time dependences of $\alpha(\omega)$. 

\subsection{Dynamical Correlations}
\label{subsec:dynamic}

\begin{figure}[]
\begin{center}
\includegraphics[width=\columnwidth,clip]{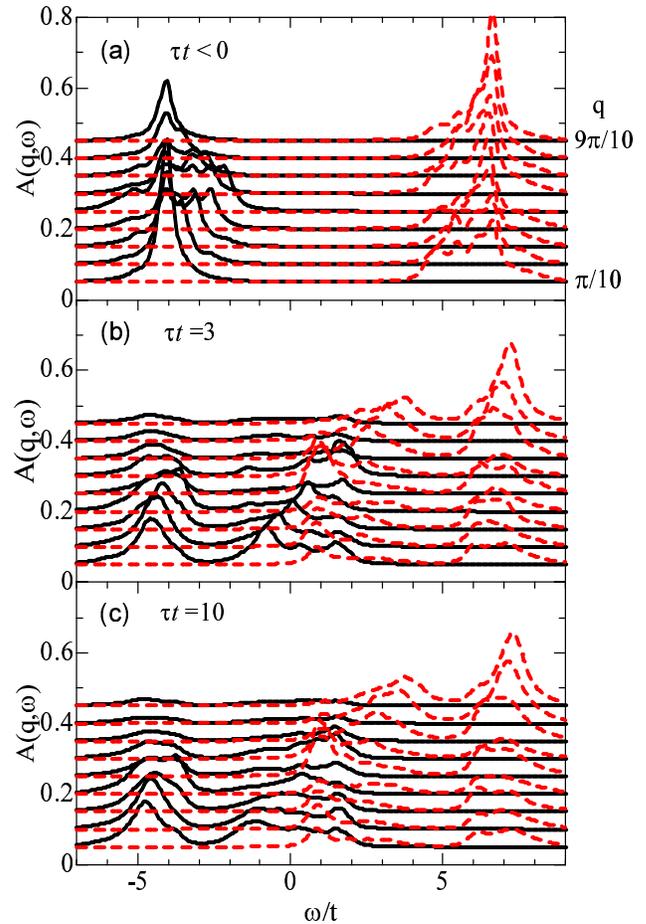}
\end{center}
\caption{(color online) The one-particle excitation spectra before pumping and those at $\tau t=$3 and 10. Solid and dotted lines are for the electron and hole parts of $A(q, \omega)$, respectively. Momentum $q$ of each panel increases from bottom to top.
}
\label{fig:Aqw_CES}
\end{figure}
%
%
The transient electronic structure in the DE model is examined directly by calculating the one-particle excitation spectra.
In Fig.~\ref{fig:Aqw_CES}, we plot $A(q, \omega)$ before and after photo-irradiation. The CO insulating gap is seen before pumping. 
The one-particle spectra are almost symmetric in the momentum space with respect to $q=\pi/2$. 
This is caused by the Brillouine-zone holding due to the staggered CO. 
There is no electron-hole symmetry in $A(q, \omega)$ because of the localized spin degree of freedom.
By photo irradiation, a photo-carrier band appears inside of the CO gap around $\omega \sim 0$, and spectral intensity of the upper (lower) band around $q=0$ ($\pi$) is weaken as shown in Fig.~\ref{fig:Aqw_CES}(b) and \ref{fig:Aqw_CES}(c).
It is noticeable that a width of the photo-carrier band becomes broad with increasing time. 

\begin{figure}[]
\begin{center}
\includegraphics[width=\columnwidth,clip]{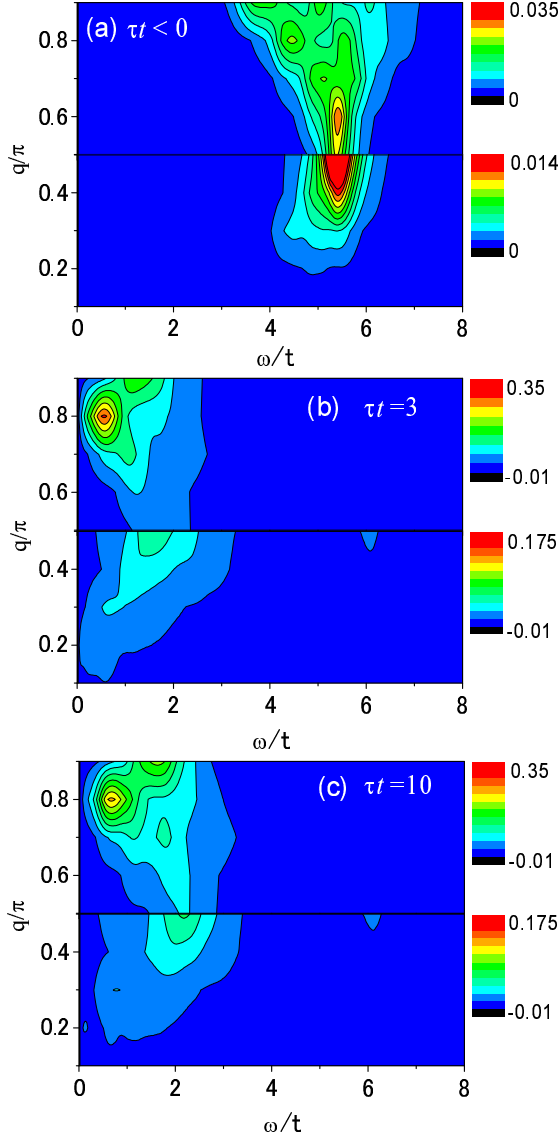}
\end{center}
\caption{(color online) 
A contour map of the imaginary part of the dynamical charge susceptibility $N(q,\omega )$ before pumping and those at $\tau t=3$ and 10.
Data in the regions of $q>\pi/2$ and $q<\pi/2$ are plotted in different scales.  
}
\label{fig:Nqw_CES}
\end{figure}
The photo-induced charge and spin dynamics are examined furthermore by calculating the dynamical charge and spin susceptibilities defined by Eq.~(\ref{172707_8Apr09}) and Eq.~(\ref{172721_8Apr09}).
Contour maps of the imaginary part of the dynamical charge susceptibility are plotted in Fig.~\ref{fig:Nqw_CES}.
The excitation spectra before pumping show 
broad continuum around $q=\pi$ and sharp peak intensity  
at $\omega/t \sim 5.5$ and $q \sim \pi/2$.  
By photo-irradiation, excitation energy significantly decreases.
Energy at the lower energy takes almost zero around $q=0$ and $\pi$. 
Large spectral weight is seen around $q=0.8\pi$,  
and total intensity is more than ten times larger than that before pumping. 
With increasing time, the band width of the continuum spectra increases weakly. 

Some features in $N(q, \omega)$ are interpreted from the kink/anti-kink picture introduced previously. 
In the $V-t$ model, which is equivalent to the $S=1/2$ XXZ model,~\cite{Ishimura_PTP} excitation energy for a kink/anti-kink pair is given by  
\begin{align}
   E(q_t, q_r) =V  -4t \cos q_t \cos q_r , 
\label{eq:deltae}
\end{align}
in the limit of $V \gg t$.
The relative and total momenta are given by $q_r=(Q_e-Q_h)/2$ and $q_t=(Q_e+Q_h)/2$, respectively, where $Q_{e(h)}$ is the momentum for the NN electron (hole) pair. 
The total momentum corresponds to $q$ in $N(q, \omega)$. 
Before photo-irradiation, charge excitations for several relative momenta $q_r$ cause excitation continuum at fixed $q(=q_t$) in $N(q, \omega)$. 
By photo-irradiation, a pair of kink/anti-kink with momenta $Q_e$ and $Q_h(=-Q_e)$ is created. In particular, in the lowest photo-excited state obtained by tuning the pump energy at the optical edge, both $Q_e$ and $Q_h$ are almost zero.  
The excitation energy in this case is $ E(0,0)=V-4t$. 
In the excitation processes for $N(q, \omega)$ in the one-photon absorbed state, ether $Q_e$ or $Q_h$ is changed, i.e. an excitation from $(Q_e, Q_h)=(0,0)$ to $(q, 0)$ or $(0, q)$. 
Energy in this state is $E(q,\pm q)=V-4t \cos^2 q$. 
Thus, the charge excitation energy in the one-photon absorbed state is given by 
\begin{align}
   \Delta E(q)&=E(q, \pm q)-E(0,0)\nonumber\\
&=2t (1-\cos 2q) .\label{132054_15Dec09} 
\end{align}
This momentum dependence explains the lower edge of the continuum
spectra after photo-irradiation shown in Figs.~\ref{fig:Nqw_CES}(b) and 9(c). 

On the other hand, continuum above the lower edge in $N(q, \omega)$ and 
increasing of the band width with increasing time are not explained only from this kink/anti-kink picture and are attributed to the localized spin degree of freedom. 
In the strong Hund-coupling limit of the DE model,  
the effective transfer integral $t_{eff}$ between the NN sites depends on the spin structure in their sites. 
It is expected that by photo-irradiation, an average value of $t_{eff}$ increases with increasing time because of reduction of the AFM correlation. This change in spin structure contributes to 
the increasing of the band width in $N(q, \omega)$. 

\begin{figure}[]
\begin{center}
    \includegraphics[width=\columnwidth,clip]{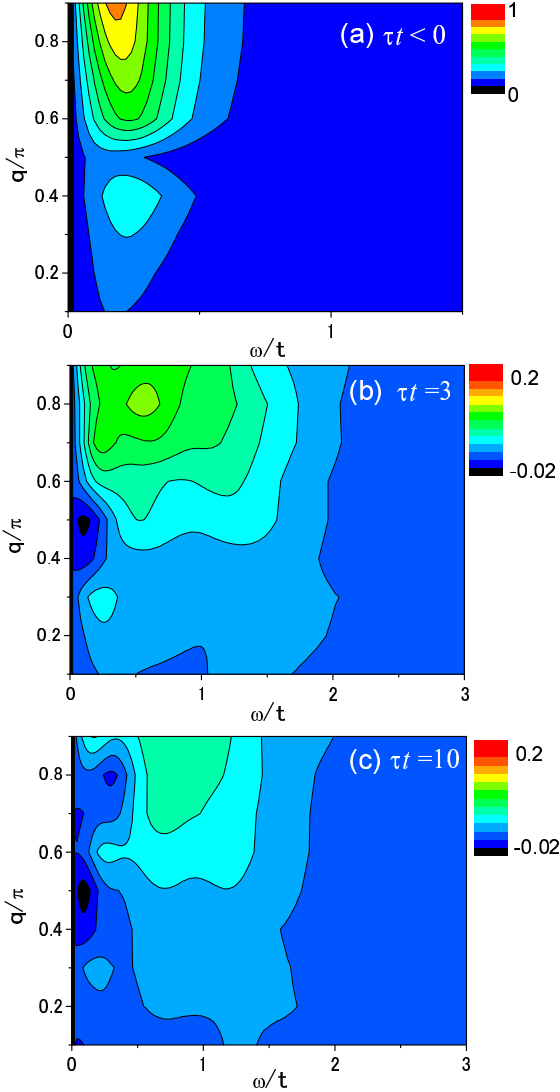}    
\end{center}
\caption{(color online) 
A contour map of the imaginary part of the dynamical spin susceptibility $S(q,\omega )$ before pumping, and those at $\tau t=3$ and 10.
}
\label{fig:Sqw_CES}
\end{figure}
Contour maps of the dynamical spin susceptibility are shown in Fig. \ref{fig:Sqw_CES}. 
Negative intensity around $\omega =0$ is due to the fact that the initial state of $S(q, \omega)$ after pumping is not the ground state but the photo-excited state where the second term in Eq.~(\ref{172721_8Apr09}) is sometime larger than the first term. 
Before pumping, the spectral weight is seen in a region of the order of $\omega \sim J_S$.
We confimred a following dispersive feature in $S(q, \omega)$ by increasing $J_S$ (not shown in the figure); 
a sine-like dispersion in the lower energy edge, continuum spectra above the edge, and a strong  intensity at $q=\pi$. 
This feature is similar to that in the de Cloiseaux-Pearson mode in the one-dimensional spin-$1/2$ AFM Heisenberg model.
After the photo-irradiation, 
the characteristic dispersion disappears and the continuum spectra are broaden 
up to $\omega \sim 2.5t$.
These changes imply a weakening of the AFM correlation by photo-irradiation 
and are consistent with the results of $S(q)$ in Fig.~\ref{fig:NqSqHsHt_CES}(b).

\subsection{Correlation between spin and charge dynamics}
\label{subsec:scaling}  
  
\begin{figure}[]
\begin{center}
\includegraphics[width=\columnwidth,clip]{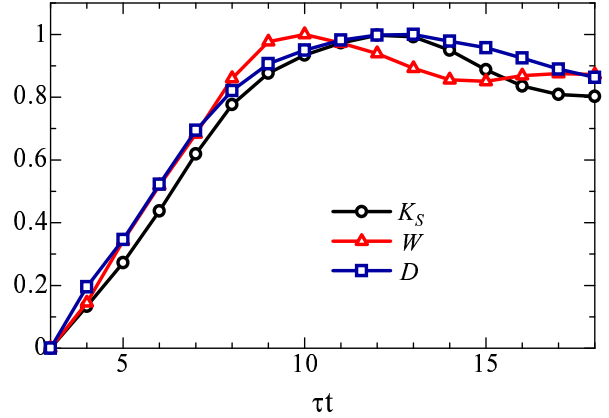}  
\end{center}
\caption{(color online) 
Time-dependences of the NN correlation between localized spins, the band width of the in-gap states in $A (q,\omega)$, and the spectral weight inside of the optical gap in $\alpha(\omega )$ denoted by $K_S$, $W$, and $D$, respectively. The data are subtracted by the data at $\tau t=3$, and normalized by differences between their minimum and maximum values.
We take $\omega_U=5t$, $\omega_L=-2.6t$, $\omega_U'=4t$ and $\omega_L'=0.1t$.
}
\label{fig:KsWD_CES}
\end{figure}
\begin{figure}[]
\begin{center}
\includegraphics[width=\columnwidth,clip]{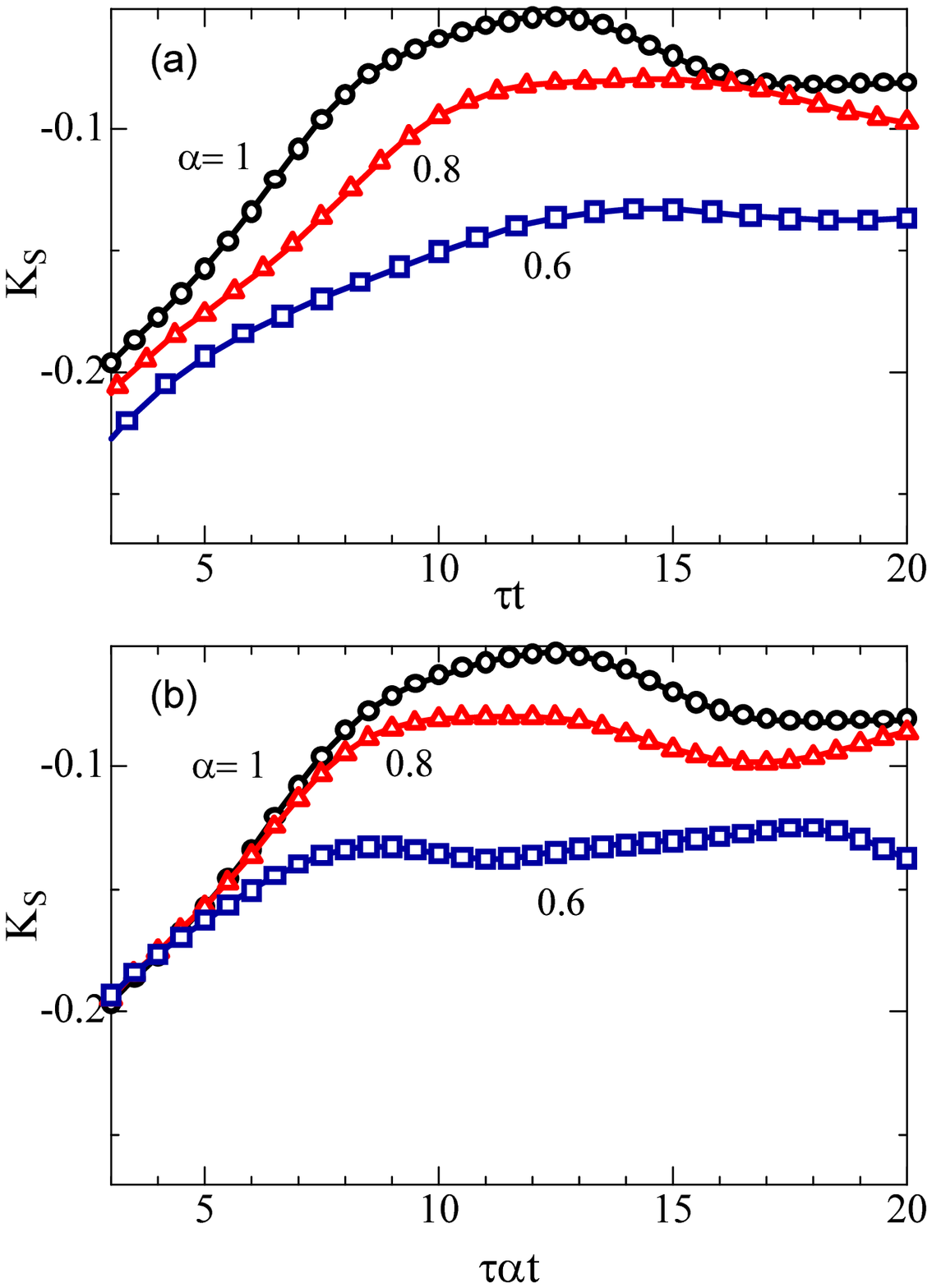}  
\end{center}
\caption{(color online)
Time-dependence of the NN spin correlation between localized spins for several value of $\alpha$. 
Horizontal axes are taken to be $\tau t$ in (a), and $\tau \alpha t$ in $(b)$.
   }
\label{fig:KsA_CES}
\end{figure}
%
In order to reveal the correlation between photo-induced charge and spin dynamics in more detail, we calculate the second moment of the in-gap band in $A(q,\omega)$ corresponding to the band width, the integrated spectral weight of the in-gap component in $\alpha (\omega)$, and the correlation function between the NN localized spins. 
These are defined by
\begin{align}
K_S=\frac{1}{L-1}\sum _{\left<ij\right>} \left< {\bm S}_i \cdot {\bm S}_{j}\right>, 
\label{eq:ks}
\end{align}
\begin{align}
W=\int  _{\omega _L } ^{\omega _U } \sum _q A(q,\omega) (\omega -\omega _c)^2d\omega, 
\end{align}
\begin{align}
D=\int_{\omega _L '} ^{\omega _U '} \alpha (\omega )d\omega , 
\end{align}
where $\omega_U$ and $\omega_{L}$ are the upper and lower edges of the in-gap component of $A(q, \omega)$, respectively,  
$\omega_{U}'$ and $\omega_{L}'$ are those of the in-gap component of $\alpha(\omega)$, 
and $\omega_c$ is the center of mass for the in-gap band defined by 
\begin{equation}
\omega _c = \frac{\int _{\omega _L} ^{\omega _U} \sum _q A(\omega ,q) \omega
 d\omega}{\int _{\omega _L} ^{\omega _U} \sum _q A(\omega ,q)d\omega} . 
\end{equation} 
Numerical values are set to be 
$\omega_U=5t$, $\omega_L=-2.6t$, $\omega_U'=4t$,
and $\omega_L'=0.1t$. 
As seen in Fig.~\ref{fig:KsWD_CES}, all three curves show similar time dependence; intensity increases linearly after pumping and saturates around $\tau t=10$.
This identical time dependence indicates a strong coupling between the charge and spin degrees of freedom in the photo-excited states.
  
The time dependence of $K_S$ at various values of $\alpha$, which is a prefactor of the transfer integral in Eq.~(\ref{171535_14Jan09}), is presented in Fig.~\ref{fig:KsA_CES}(a). The results are also plotted as a function of the time scaled by the transfer integral, $\tau \alpha t$, in Fig.~\ref{fig:KsA_CES}(b). 
The slope of the curve increases with decreasing $\alpha$. 
By reploting as a function of time scaled by $\alpha t$, 
the slopes for several value of $\alpha $ are almost identical in a region of $\tau t = 3-7$. Through this analysis, we conclude that photo induced charge and spin dynamics are controlled by the electron transfer integral for the conduction electrons. 

\section{Pump-photon density dependence}
\label{sect:amp}
So far, we restrict our calculations to the one-photon absorbed states in a finite size system. In this section, we introduce pump-photon density dependence of the charge and spin dynamics in the extended DE model. 
Let us go back to the formulation for the electron-photon interaction. 
We consider a condition that amplitude of the pump photon is not weak and the interaction Hamiltonian between electron and photon in Eq.~(\ref{eq:ha}) 
is not treated perturbatively. 
For convenience in numerical calculations, 
instead of Eq.~(\ref{eq:Apump}), we introduce a Gaussian-type of the pump-photon vector potential given by 
\begin{align}
  A_{\rm pump}(\tau)=A_1 \exp \left[ -\frac{\gamma _0 ^2(\tau-\tau_0)^2}{2}\right]\cos \omega _0 (\tau-\tau_0) , 
\label{eq:newpump}
\end{align}
where $\tau_0(>0)$ is a center of the wave packet. 
We have checked that the results, such as the spin correlation function, calculated with this type of pump-photon vector potential qualitatively reproduce the results with the Loretzian-type vector potential introduced in Eq.~(\ref{eq:Apump}). 
Time evolution of the system is governed by a sum of the electronic part and the interaction part, i.e. ${\cal H}(\tau)={\cal H}_0+{\cal H}'(\tau)$ where ${\cal H}_0$ and ${\cal H}'(\tau)$ are given in Eq.~(\ref{171535_14Jan09}) and Eq.~(\ref{eq:ha}), respectively, with Eq.~(\ref{eq:newpump}). 
We assume that $\tau_0 \gg \gamma_0^{-1}$, and 
the system at time $\tau=0$ is in the ground state of ${\cal H}_0$, i.e. $|\phi(\tau=0) \rangle=|0 \rangle$. 
Time evolution of the wave function is calculated by 
Eq.~(\ref{184408_9Nov09}) from $|\phi(\tau=0)\rangle$. 
In the Lanczos method, we confirmed the parameters $M_2=20$ and $\delta \tau =0.01/t$ are sufficient 
to obtain the reliable results. 
In the multi-target DMRG algorithm, we adopt
$\left\{\ket{\phi(\tau )},\ket{\phi(\tau +d\tau )} \right\}$ as the target states 
in the mixed-state density matrix defined in Eq.~(\ref{eq:dm}). 
We chose the parameters $M_2=20$ and $\delta \tau=0.1/t$ in Eq.~(\ref{184408_9Nov09}), 
and the truncation number $m=600 - 800$. 
In Eq.~(\ref{eq:newpump}), the pump-photon energy is chosen to be $\omega_0=5t$ which is around a middle of the lowest absorption band shown in Fig.~\ref{fig:gsNSaw}. 
We chose the damping factor $\gamma_0=2t$ in order to demonstrate clearly the pump-photon density dependence. 
System size is chosen to be $L=$9, 13 and 17. 

\begin{figure}[t]
\begin{center}  
  \includegraphics[width=\columnwidth,clip]{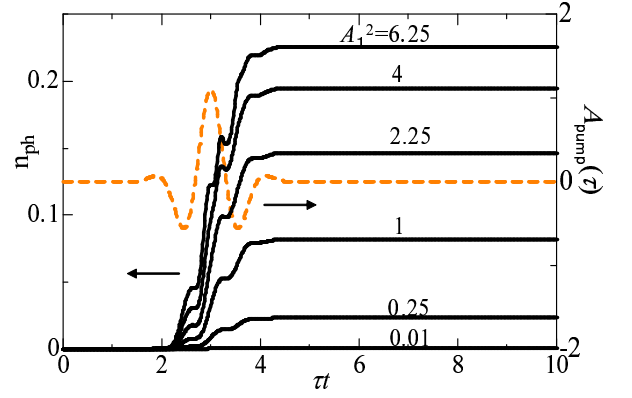}
\end{center}
\caption{(color online)  
Time dependence of the absorbed photon density $n_{ph}(\tau)$ for
 several value of $A_1^2$ (solid lines), and that of the pump-photon
 vector potential $A_{\rm pump}(\tau)$ in the case of $A_1 ^2 =1$ (broken line).}
\label{fig:amp_a0}
\end{figure}
\begin{figure}[t]
\begin{center}
  \includegraphics[width=\columnwidth,clip]{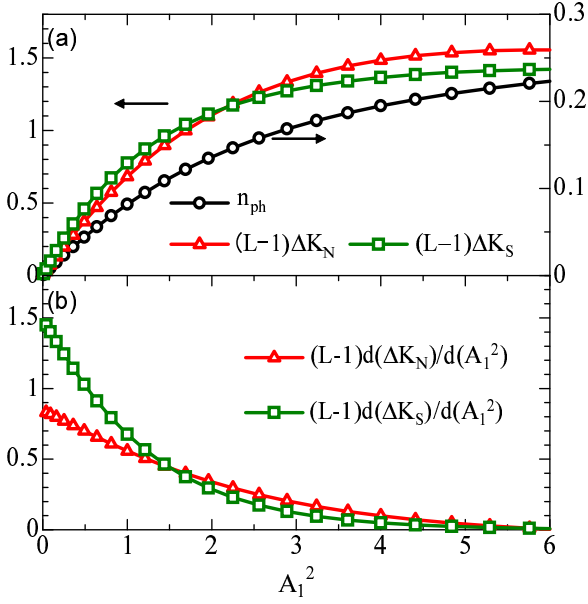}
\end{center}
\caption{(color online) 
(a) The absorbed photon density $n_{ph}(\tau)$, the charge correlation function $(L-1)\Delta K_N$, and the spin correlation function $(L-1)\Delta K_S$ as functions of the pump-photon amplitude. 
(b) The first derivative of $K_N$ and $K_S$ with respect to $A_1^2$, i.e. 
$(L-1)d(\Delta K_N)/d( A_1^2)$ and $(L-1)d(\Delta K_S)/d(A_1^2)$.
Time is chosen to be $t\tau =5$.
}
\label{fig:a0-dep}
\end{figure}
In Fig.~\ref{fig:amp_a0}(a), the photon densities absorbed in the system are plotted for several values of a prefactor $A_1^2$ in the pump-photon vector potential in Eq.~(\ref{eq:newpump}). 
We define the absorbed photon density at time $\tau$ by 
\begin{equation}
n_{ph}(\tau)=\frac{\langle {\cal H} (\tau) \rangle -\langle {\cal H} (- \infty) \rangle}{L \omega_0} , 
\label{eq:density}
\end{equation}
where ${\cal H}(-\infty)$ is the Hamiltonian at time $\tau \ll \tau_0$. 
We also plot $A_{\rm pump}(\tau)$ in the same figure. 
The absorbed photon density increases abruptly around $\tau=\tau_0=3/t$, and is saturated around $\tau=5/t$. A saturated value of $n_{ph}$ monotonically increases with increasing $A_1^2$. 

We focus on the electronic states at $\tau=5/t$ where $A_{\rm pump}(\tau)$ is almost fully damped 
and $n_{\rm ph}(\tau)$ is saturated. 
The following results are not sensitive to a choice of a value of $\tau$. 
In Fig.~\ref{fig:a0-dep}(a), the absorbed photon density and the charge and spin correlation functions at time $\tau=5/t$ are plotted as functions of the pump-photon intensity. 
We define that $\Delta K_{N(S)}$ is a difference of $K_{S(N)}$ at $\tau=5/t$ from that at $\tau \ll \tau_0$. 
The spin correlation function $K_S$ was introduced in Eq.~(\ref{eq:ks})
and the charge correlation function is defined by  
\begin{align}
K_N=\frac{1}{L-1}\sum_{\langle ij \rangle} \langle n_i n_j \rangle . 
\end{align}
Both the two increase with increasing $A_1^2$ and tend to be saturated around $A_1^2=3-4$. 
In order to compare the spin and charge dynamics in more detail, we plot the first derivatives of $\Delta K_S$ and $\Delta K_N$ with respect to $A_1^2$ in Fig.~\ref{fig:a0-dep}(b). 
It is shown that a change in $\Delta K_S$ is more sensitive than that in $\Delta K_N$ 
in weak photon density, and is almost saturated around $A_1^2=3$. 

\begin{figure}[]
\begin{center}
    \includegraphics[width=\columnwidth,clip]{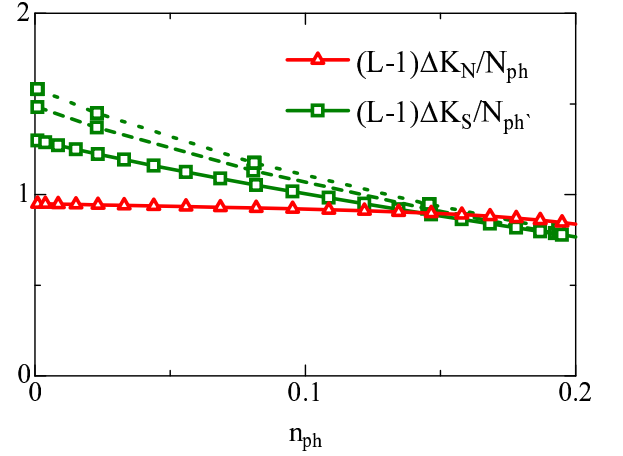}
\end{center}
\caption{(color online) 
Differences in the charge and spin correlation functions devided by absorbed photon number. 
Data are plotted as functions of the absorbed photon density. 
Solid, broken and dotted lines are for the results in system sizes $L=$9, 13 and 17, respectively. 
}
\label{fig:nph-dep}
\end{figure}
To clarify this different spin and charge dynamics furthermore, we plot changes in the spin and charge correlation functions divided by the absorbed photon number $N_{ph}$ as functions of $n_{ph}$ in Fig.~\ref{fig:nph-dep}. 
We introduce 
\begin{align}
N_{ph}=L n_{ph} . 
\end{align}
Values of $\Delta K_N/N_{ph}$ are almost independent of $n_{ph}$ 
and are identical to be about one. This result does not show remarkable size dependence. 
On the other hand, $\Delta K_S/N_{ph}$ shows clear $n_{ph}$ dependence and almost linearly decreases with increasing $n_{ph}$. A weak system size dependence is seen in $\Delta K_S/N_{ph}$ in a region of small $n_{ph}$. 

This difference in the spin and charge dynamics is interpreted as follows. 
The result of $\Delta K_N/N_{ph} \sim 1$ implies that one-photon generates one electron-hole pair. This is explained by the kink/anti-kink picture in a one-dimensional chain introduced in the previous section. On the contrary, it is shown that destruction of the AFM spin correlation by pumping is effective in the weak photon density case. 
This may be attributed to the interaction between the photo-carriers. 
In the case of large $V$, the photo-carriers are not exchanged with each other in a chain. 
Each carrier destroys the AFM spin correlation through the DE interaction.
In the case of high photo-carrier density, i.e. large $n_{ph}$, regions where photo-carrier destroys the AFM correlation are overlapped and destructions of the AFM correlation are interrupted with each other. 
%

\begin{figure}[]
\begin{center}
   \includegraphics[width=\columnwidth,clip]{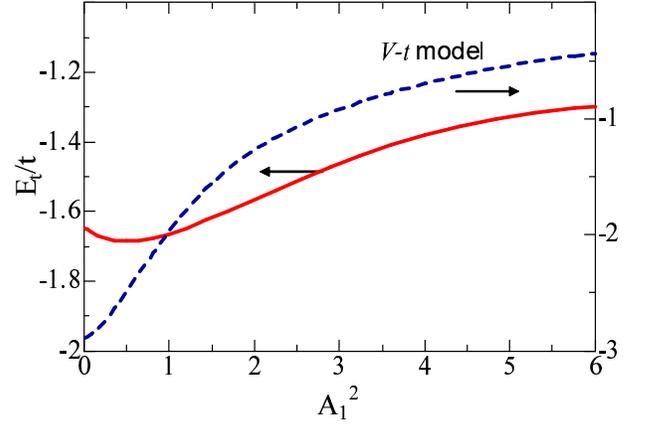}
\end{center}
\caption{(color online) 
The kinetic energy as functions of the pump-photon intensity. 
Broken line is for the results in the spin-less $V-t$ model.  
}
\label{fig:kin-a0-dep}
\end{figure}
Roles of spin degree of freedom in photo-induced dynamics are also seen in the photon-density dependence of the kinetic energy. 
In Fig.~\ref{fig:kin-a0-dep}, the kinetic energies in the extended DE model and the $V-t$ model are plotted as functions of the pump-photon intensity. 
In the calculation of the $V-t$ model, 
one dimensional cluster of $L=9$ and $N=5$ with the open-boundary condition is used. 
The parameter values are chosen to be $V=5t$, $\omega_0=5t$ and $\gamma_0=2t$. 
We plot $E_t$ defined in Eq.~(\ref{eq:et}) and its correspondence in the $V-t$ model at $\tau=5/t$ where $n_{ph}$ is saturated. 
A non-monotonic dependence on $A_1^2$ is seen in the DE model; 
$E_t$ decreases at first in a region of weak photon density, and increases above $A_1^2 \sim 0.5$. 
This is in contrast to the results in the $V-t$ model where $E_t$ monotonically increases with the photon amplitude. 
That is, the spin degree of freedom qualitatively changes the photo-induced dynamics in a weak intensity region. 
This observation is explained from a competition between carrier excitation by photo irradiation and the cooperative weakening of the charge correlation and the AFM correlation.  
A monotonic increasing of the kinetic energy is caused by 
a carrier excitation by pump photon.  
On the contrary, the charge correlation is reduced by a weakening of the AFM correlation due to photo-carrier motion. 
In a region of small $A_1^2$, the latter overcomes the former, 
and, and as a result, $E_t$ decreases with $A_1^2$. 
On the other side, in the strong pump-photon density, a destruction of AFM correlation is saturated, and the photo-carrier excitation effect emerges.

\section{Discussion and Concluding Remarks}
\label{sect:dis}

In this section, we summarize the present numerical results and discuss implications for the photo-induced phenomena in perovskite manganites. 
The present main results are listed: 
1) a new state appears inside of the insulating gap by photo-irradiation (see Figs.~\ref{fig:gsNSaw} and \ref{fig:Aqw_CES}).  2) time dependence of the NN spin correlation $K_S$ is strongly correlated with those of the band width $W$ of the in-gap states in $A(q,\omega)$ and the spectral weight $D$ inside of the optical gap in $\alpha(\omega)$. 
These are scaled by a universal curve (see Fig.~\ref{fig:KsWD_CES}). 
3) The characteristic time scale $\tau_S$, when a change in $K_S$ is saturated, is governed by the electron transfer $t$ (see Fig.~\ref{fig:KsA_CES}), i.e. $\tau_S \sim 10/t$. 
Although the present our calculations have been carried out in one-dimensional finite-size clusters, the characteristics in the photo-induced dynamics listed above are qualitatively reproduced by the time-dependent unrestricted Hartree-Fock method in two-dimensional systems.~\cite{Kanamori_PRL} 
By using the realistic parameter values for manganites, this time scale $\tau_S$ is about few fs. The present calculations predict that both the CO correlation and 
the short-range AFM correlations collapse cooperatively within this time scale and a metallic state appears in the optical spectra and the photoemission spectra. 
This ultra-fast CO melting and generation of a metallic state are consistent with the recent pump-probe experiments in manganite; a melting of charge order within about 30fs by irradiation was observed.~\cite{Matsuzaki_PRB79} 

We note that the weakening of the NN spin correlation within few fs is not contradict to the magneto-optical Kerr spectroscopy measurement. 
In several charge ordered manganites, it was shown that the photo-induced spin dynamics observed by the Kerr spectroscopy is slower than the photo-induced charge dynamics probed by the optical reflection.~\cite{Miyasaka_PRB74,Matsubara_PRL99,McGill_PRL93,Matsubara_JPSJ78,Mertelj_EPL86}  
These experiments and the present theory suggest that the short-range and long-range spin dynamics are governed by the different mechanisms, the electron transfer and the spin-orbit coupling, respectively. 
The macroscopic spin dynamics cannot be dealt with by the present ED method where the spin angular moment is conserved in time evolution. 
This issue was studied in detail in our previous paper~\cite{Kanamori_PRL} by utilizing the time-dependent Hartree-Fock method with spin relaxation processes. 
Photo-induced dynamics in the macroscopic magnetic moment is controlled by the phenomenological spin relaxation constant $\Gamma$, and the finite magnetic moment appears in the longer time scales of $\tau_L \sim 1/\Gamma$ than $\tau_S$. 
One noticeable point is that change in the charge sector from the CO to the metal is almost completed within the short time scale $\tau_S$, and growing of the macroscopic magnetic moment around $\tau_L$ does not associated with change in both the short- and long-range charge correlations. 
This is attributed to the fact that the spin-orbit interaction, by which 
the conservation of the macroscopic spin angular moment is only broken, play a crucial role on the photo-induced macroscopic spin dynamics. 

It is shown that, even in the short-time scale, photo-induced spin and charge dynamics does not always show identical behaviors; the change in the short-range spin sector becomes remarkable in the case of weak pump-photon density, in comparison with that in the charge sector. This is seen in the spin correlation function and the kinetic energy as functions of the pump-photon density (see Figs.~\ref{fig:a0-dep} and \ref{fig:kin-a0-dep}). These different dynamics between charge and spin are caused by the fact that the initial CO order is collapsed directly by the photon, while the AF spin correlation is collapsed by carrier motion. The detailed measurements of the total weight of the optical conductivity spectra may 
detect the non-monotonic behavior of the kinetic energy as a function of pump-photon density. 

In conclusion, we study numerically the photo-induced electronic dynamics in the generalized DE model in one-dimensional chains. 
The Lanczos and DMRG methods are utilized to calculate the time evolution after photo-irradiation and the several spectral functions. 
Roles of the localized-spin degrees of freedom in the DE model are focused on. 
By pump-photon irradiation into the CO and AFM insulating state, 
the collapsing of the AFM correlation and the appearance of a metallic state occur cooperatively. 
This time evolution is governed by the electron transfer integral of the conduction electron. 
The numerical results are explained by the charge kink/anti-kink picture and their spin-dependent kinetic motion. 
The pump-photon density dependence of the spin and charge dynamics are also examined. 
Roles of the spin degree of freedom are remarkable in the case of the weak pump-photon density.  

\appendix
\section{V-t model in one dimension}
\label{appendixa}

In this APPENDIX, the effective model obtained from the $V-t$ model 
and an analytical formulation for the optical absorption spectra in this model are presented. 
We start from the one-dimensional spin-less $V-t$ model introduced in Eq.~(\ref{eq:vt})
where the numbers of sites and electrons are $L$ and $L/2$, respectively. 
In the limit of $V \gg t$, the ground state electron configurations are denoted as $\cdots 101010101 \cdots$ where 1 and 0 represent electron occupied and unoccupied sites, respectively. 
The low energy excited states are given by the configurations where one NN electron pair and one NN hole pair exist shown as $\cdots 100101101 \cdots $. 
In this subspace, the effective Hamiltonian in the limit of $V \gg t$ is obtained by the canonical perturbational expansion. 
In the case where the electron pair locates in the left side of the hole pair in a chain, 
the Hamiltonian is given as 
\begin{align}
\widetilde {\cal H}_{Vt}=-t \sum_{i}^{L/2} \left ( a_i^\dagger a_i+b_i^\dagger b_{i+1} +h.c. \right ) . 
\label{eq:effective}
\end{align}
We introduce the two fermion operators,  
\begin{align}
a_i=n_{2i}d_{2i-1}+n_{2i-1}d_{2i},  
\label{eq:aa}
\end{align}
\begin{align}
b_i=(1-n_{2i})d_{2i-1}^\dagger-(1-n_{2i-1})d_{2i}^\dagger , 
\label{eq:ab}
\end{align}
which correspond to the annihilation operator of the
NN electron pair and that of the NN hole pair, respectively.
We impose the relations 
\begin{align}
\sum_i a_i^\dagger a_i=\sum_i b_i^\dagger b_i=1 , 
\end{align}
and we have the following anti-commutation relations 
\begin{align}
\{ a_i^\dagger , a_i \}&=\delta_{ij} \left ( n_{2i}+n_{2i+1} \right ),   \\
\{ b_i^\dagger , b_j \}&=\delta_{ij} \left [
\left ( 1-n_{2i} \right) + \left (1-n_{2i+1} \right ) \right ], 
\end{align}
and other anti-comutators are zero. 
The operators in Eq.~(\ref{eq:aa}) and Eq.~(\ref{eq:ab}) satisfy the conditions
\begin{align}
a_i^\dagger b_i^\dagger=a_i b_i^\dagger=b_i a_i^\dagger=0. 
\end{align}

In this model with the open-boundary condition, 
the eigen states are classified by two momenta $k_1$ and $k_2$ for a electron pair and a hole pair as 
\begin{align}
 \ket{k_1, k_2}=&\frac{2}{L+1} \sum _{i>j} 
\Bigl( \sin x_i k_1 \sin x_j k_2 
-\sin x_i k_2 \sin x_j k_1 \Bigr) \nonumber \\
&\times a_i ^\dag b_j ^\dag \ket{0} .
\end{align}
The corresponding eigen value is 
\begin{align}
E_{k_1, k_2} = -2t \left ( \cos k_1 +\cos k_2 \right ). 
\end{align}
We introduce the momentum $k=n \pi/(L+1)$ and an integer number $n ( = 1,2,\cdots, L)$.
In the present scheme, the current operator is given by
\begin{align}
j=-it \sum _i \left( a_i ^\dag a_{i+1}-b_i ^\dag b_{i+1} -H.c. \right ).
\end{align}
A matrix element of $j$ between two states is obtained by
\begin{align}
& \bra{k_{f1} k_{f2} }j\ket{k_{i1} k_{i2}} 
= \left( \frac{2}{L}  \right)^2  it  \int _ 0 ^L d x_i \int _0 ^{x_i} dx_j \nonumber \\
&\times \left ( \sin x_i k_{f1} \sin x_j k_{f2} 
 -\sin x_i k_{f2} \sin x_j k_{f1}  \right ) \nonumber \\
& \times \Bigl [ 2\sin k_{i1} 
\left ( \cos x_i k_{i1} \sin x_j k_{i2} +\cos x_i k_{i2} \sin x_j k_{i1} \right ) 
\nonumber \\
&\  - 2\sin k_{i2} 
\left ( \sin x_i k_{i1} \cos x_j k_{i2} +\sin x_i k_{i2} \cos x_j k_{i1} \right ) 
\Bigr ] .
\label{eq:ap1}
\end{align}
In the thermodynamic limit $L \to \infty$, the total momentum $ k^{\rm tot}(=k_1 - k_2)$ is zero in the optical process, and the eigen states are characterized by the relative momentum $q[ \equiv (k_1 + k_2)/2]$. 
By carrying out the integrals in Eq.~(\ref{eq:ap1}), the transition probability from the state with $q_i$ to that with $q_f$, i.e. 
$I(q_f, q_i) \equiv |\langle q_f |j| q_i \rangle |^2$, is given by 
\begin{align}
I(q_f, q_i)= \frac{6^2 t^2}{\pi ^2} \sin ^2 q_i, 
\end{align}
for the minimum momentum change $q_f=q_i+\pi/L$, 
and 
\begin{align}
I(q_i, q_f)=\frac{8^2 t^2}{ L^2}  \left ( \frac{q_f \sin q_i }{ q_f^2-q_i^2} \right )^2  , 
\end{align}
for others. 
Above two are summarized in Eq.~(\ref{eq:pro}).

\begin{acknowledgments}
The authors would like to thank 
K. Nasu, K. Miyano, H. Okamoto, S. Koshihara, S.~Iwai, T. Arima and J. Ohara for their valuable discussions. 
This work was supported by JSPS KAKENHI (21540312, 21224008, 21740268), CREST, 
Tohoku University "Evolution" program, 
and Grand Challenges in Next-Generation Integrated Nanoscience.
YK is supported by the global COE program "Weaving Science Web beyond Particle-Matter Hierarchy" of MEXT, Japan.
\end{acknowledgments}

\newpage

\end{document}